\documentclass[aps,twocolumn,amsmath,letterpaper]{revtex4}
\usepackage{amssymb}
\usepackage{graphicx}
\usepackage{array}
\usepackage{hhline}
\usepackage{longtable}

\renewcommand{\thetable}{\Roman{table}} \thetable

\begin{document}

\title{Reentrant and Forward Phase Diagrams of

the Anisotropic Three-Dimensional Ising Spin Glass}
\author{Can G\"uven$^1$, A. Nihat Berker$^{1-3}$, Michael Hinczewski$^3$, and Hidetoshi Nishimori$^4$}
\affiliation{$^1$Department of Physics, Ko\c{c} University, Sar\i
yer 34450, Istanbul, Turkey,} \affiliation{$^2$Department of
Physics, Massachusetts Institute of Technology, Cambridge,
Massachusetts 02139, U.S.A.,} \affiliation{$^3$Feza G\"ursey
Research Institute, T\"UBITAK - Bosphorus University,
\c{C}engelk\"oy 34680, Istanbul, Turkey,}
\affiliation{$^4$Department of Physics, Tokyo Institute of
Technology, Oh-okayama, Meguro-ku, Tokyo 152-8551, Japan}

\begin{abstract}  The spatially uniaxially anisotropic $d=3$ Ising spin glass is
solved exactly on a hierarchical lattice.  Five different ordered
phases, namely ferromagnetic, columnar, layered, antiferromagnetic,
and spin-glass phases, are found in the global phase diagram.  The
spin-glass phase is more extensive when randomness is introduced
within the planes than when it is introduced in lines along one
direction.  Phase diagram cross-sections, with no Nishimori
symmetry, with Nishimori symmetry lines, or entirely imbedded into
Nishimori symmetry, are studied.  The boundary between the
ferromagnetic and spin-glass phases can be either reentrant or
forward, that is either receding from or penetrating into the
spin-glass phase, as temperature is lowered.  However, this boundary
is always reentrant when the multicritical point terminating it is
on the Nishimori symmetry line.

\end{abstract}
\maketitle
\def\s{\rule{0in}{0.28in}}
\setlength{\LTcapwidth}{\columnwidth}

\begin{figure*}
\includegraphics*[scale=1]{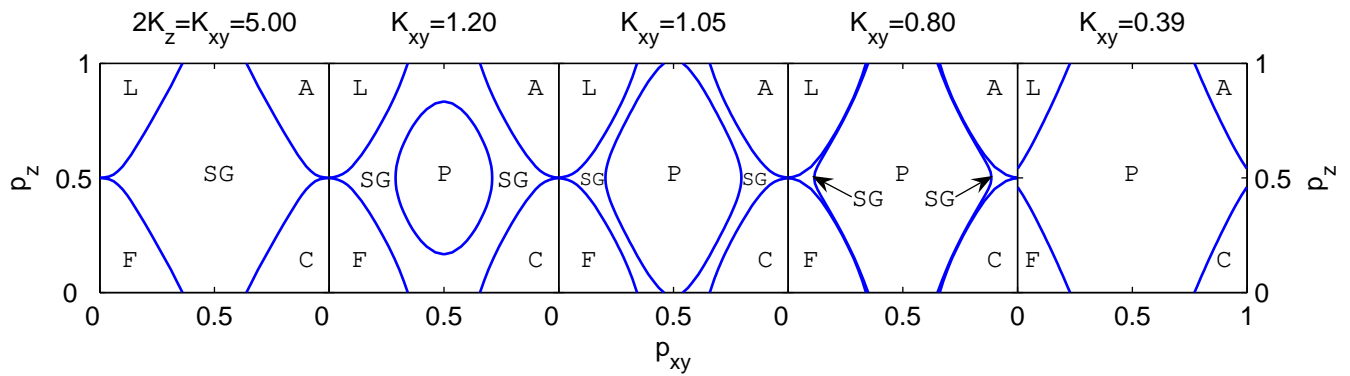}
\caption{(Color on-line) Constant-temperature cross-sections of the
global phase diagram for $K^z/K^{xy}=0.5$, as a function of $p_{xy}$
and $p_z$, which are the concentrations of antiferromagnetic $xy$
and $z$ bonds, respectively. At low temperatures (high $K^{xy}$),
the central spin-glass (SG) phase separates the corner ferromagnetic
(F), columnar (C), antiferromagnetic (A), and layered (L) phases.
The diagrams are twofold symmetric along each axis, but not fourfold
symmetric, due to the difference between longitudinal ($p_{xy}=0$)
and transverse ($p_z=0$) spin glasses. As temperature increases, the
paramagnetic (P) phase appears at the central point, first reaches
the transverse spin-glass system and eliminates the spin-glass
phase, then reaches the longitudinal spin-glass system and
eliminates the spin-glass phase.  In the latter system, the
spin-glass and paramagnetic phases simultaneously occur for a very
narrow range of temperatures, as also seen in the inset in the lower
left panel of Fig.3.}\label{lfig1}
\end{figure*}

\section{Introduction}

The Ising spin glass \cite{NishimoriBook} yields a phase diagram
with a distinctively complex ordered phase, in $d=3$. A wide
accumulation of methods and results has occurred for this system.
Most remarkably, in spite of its high spatial dimension and complex
ordering behavior, exact or precise information is being obtained
for this system.\cite{NishimoriA, NishimoriB, NishimoriC,
NishimoriD, NishimoriE, NishimoriNemoto, Maillard, Takeda2, Takeda1,
HinczBerker1, Nishimori}  Thus, in the phase diagram in terms of
temperature and concentration of antiferromagnetic bonds, the
occurrence of the Nishimori symmetry line has been deduced
\cite{NishimoriA, NishimoriB} and the accurate location of the
multicritical point has been predicted \cite{Takeda1, Nishimori}.
Furthermore, in systems with the Nishimori symmetry, it has been
shown that the ferromagnetic phase cannot extend to
antiferromagnetic bond concentrations beyond that of the
multicritical point.\cite{NishimoriA, NishimoriB}  The two remaining
options being a straight line or a reentrance situation, subsequent
works \cite{Migliorini, Nobre} on hierarchical lattices have shown
that for these systems, the spin-glass phase diagram is reentrant,
namely that below the multicritical point, the ferromagnetic phase
recedes from the spin-glass phase as temperature is lowered.  Exact
results recently have also been extended to Potts spin
glasses.\cite{Ohzeki}  These results complement recent precise
calculations, using Monte Carlo simulations, on cubic
lattices.\cite{Binder1, Binder2, Binder3, Katzgraber, Hasenbusch,
Young}

A spatially uniaxially anisotropic $d=3$ system is studied in this
work, to our knowledge the first study of quenched randomness and
frustration in a spatially anisotropic higher-dimensional system. In
fact, both anisotropy and quenched randomness have acquired
increased relevance from high-temperature superconductivity
results.\cite{HinczBerker2, HinczBerker3}  Our calculation is exact
for a hierarchical lattice and approximate for a cubic lattice.  We
find a rich phase diagram (e.g., Fig.\ref{lfig1}) with five
different ordered phases, namely with ferromagnetic,
antiferromagnetic, layered, columnar, and spin-glass order.  The
spin-glass phase is more extensive when randomness is introduced
within the planes than when it is introduced in lines along one
direction.

The global phase diagram includes cross-sections with no Nishimori
symmetry, cross-sections with Nishimori symmetry lines, and a
cross-section entirely imbedded within Nishimori symmetry. Thus, the
multicritical point between the spin-glass, ferromagnetic, and
paramagnetic phases, previously found to occur on the Nishimori
symmetry line, is also found here at points with no Nishimori
symmetry, but renormalizes to a fixed distribution of interaction
probabilities that obeys Nishimori symmetry.  Nevertheless, we find
that the boundary between the ferromagnetic and spin-glass phases
can be either reentrant or forward, that is either receding from or
penetrating into the spin-glass phase, as temperature is lowered.
When the multicritical point is not on the Nishimori symmetry line,
the ferromagnetic-spin glass boundary can be reentrant or forward.
However, when the multicritical point is on the Nishimori symmetry
line, this boundary is always reentrant \cite{Migliorini, Nobre},
consistently with the rigorous result \cite{NishimoriA, NishimoriB}.

\begin{figure}[h]
\includegraphics*[scale=1]{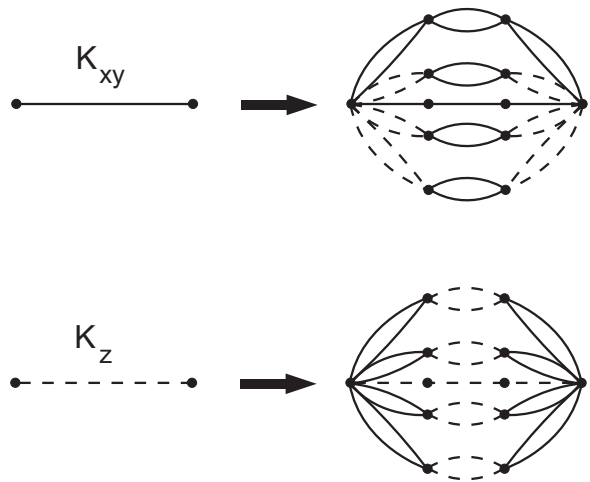}
\caption{Construction of the uniaxially anisotropic d=3 hierarchical
model.  Two graphs are mutually and repeatedly self-imbedded. Note
that for $K^{xy} = 0, K^z = 0$, and $K^{xy} = K^z$, the system
reduces respectively to the $d=1$, isotropic $d=2$ and $d=3$
systems.} \label{lfig2}
\end{figure}

\section{Uniaxially Anisotropic Spin Glass}

The uniaxially anisotropic Ising spin-glass system has the
Hamiltonian
\begin{equation}\label{eq:1}
-\beta H = \sum_u{\sum_{\langle ij \rangle_u} K^u_{ij} s_i s_j}\,,
\end{equation}
where $s_i = \pm 1$  at each site $i$, $\langle i j \rangle_u$
denotes a sum over nearest-neighbor pairs of sites along the $z$
direction ($u=z$) or in the $xy$ plane ($u=xy$), and the bond
strengths $K^u_{ij}$ are equal to $K^u>0$ with probability $1-p_u$
and $-K^u$ with probability $p_u$, respectively corresponding to
ferromagnetic and antiferromagnetic interaction. When imbedded into
a cubic lattice, the Hamiltonian (\ref{eq:1}) yields a uniaxially
anisotropic $d=3$ system.

Hierarchical lattices are $d$-dimensional lattices yielding exact
renormalization-group solutions to complex statistical mechanics
problems.  These lattices are constructed by the repeated
self-imbedding of a graph into a bond \cite{BerkerOstlund, Kaufman,
Kaufman2}.  The shortest path between the external vertices of the
graph gives the length rescaling factor $b$ and the number of bonds
in the graph gives the volume rescaling factor $b^d$, from which the
dimension $d$ is determined. Hierarchical lattices have been used to
study a wide variety of problems, including chaotic rescaling
\cite{McKay, Garel}, spin-glass \cite{Migliorini}, random-field
\cite{Falicov2}, Schr\"odinger equation \cite{Domany},
lattice-vibration \cite{Langlois}, dynamic scaling
\cite{Stinchcombe}, random-resistor network \cite{Medina}, aperiodic
magnet \cite{Haddad}, complex phase diagram \cite{Le}, directed-path
\cite{Derrida, daSilveira}, heteropolymer \cite{Tang},
directed-polymer \cite{Garel2}, and, most recently, scale-free and
small-world network \cite{Hinczewski2, Hinczewski3, ZRZ, ZZZ, ZZC,
Ben-Avraham, Rozenfeld, Khajeh} systems, etc. More recently,
hierarchical lattices have been created \cite{Erbas} for the study
of spatially anisotropic systems. The mutual repeated self-imbedding
of two appropriately chosen graphs, with differentiated
interactions, yields a uniaxially anisotropic system, whereas a
higher number of graphs is needed to achieve higher spatial
anisotropy.\cite{Erbas} These hierarchical systems must reduce to
isotropy and/or lower spatial dimensions when corresponding
interactions are set equal to each other or to zero, as illustrated
in Fig.\ref{lfig2}.  An anisotropic hierarchical lattice has already
been used to obtain the phase diagram of the uniaxially anisotropic
$d=3$ $tJ$ model of electronic conduction.\cite{HinczBerker2}  When
imbedded into the hierarchical lattice of Fig.\ref{lfig2}, the
Hamiltonian (\ref{eq:1}) yields a uniaxially anisotropic $d=3$
spin-glass system that is exactly soluble.

\section{Exact Renormalization-Group Solution: Flows of the
Quenched Distributions of the Anisotropic Spin-Glass Interactions}

The renormalization-group solution proceeds in the direction
opposite to the construction of a hierarchical model.  Each graph is
replaced by a renormalized bond via summation over the spins on the
internal sites of the graph.  This is achieved by a combination of
two types of steps: the replacement, by a single bond
$\widetilde{K}_{ij}$, of two bonds that are either in parallel,
referred to as bond-moving:

\begin{equation}\label{eq:2}
\widetilde{K}_{ij} =  K_{ij}^I + K_{ij}^{II},
\end{equation}

\noindent or in series, referred to as decimation:

\begin{equation}\label{eq:3}
\widetilde{K}_{ik} = \frac{1}{2} \ln \left[
\frac{\cosh(K_{ij}+K_{jk})}{\cosh(K_{ij}-K_{jk})} \right]\,.
\end{equation}

\noindent The quenched probability distribution $\widetilde{{\cal
P}}(\widetilde{K})$ of the replacing bond is calculated by the
convolution

\begin{equation}\label{eq:4}
{\widetilde{{\cal P}}}(\widetilde{K}) = \int dK^I dK^{II} {\cal
P}_I(K^I){\cal P}_{II}(K^{II}) \delta(\widetilde{K} -
R(K^I,K^{II}))\,,
\end{equation}

\noindent where $R(K^I,K^{II})$ is the right-hand side of
Eq.(\ref{eq:2}) or (\ref{eq:3}), $K^I$ and $K^{II}$ are the
interactions entering the right-hand side of either of these
equations, with quenched probability distributions ${\cal P}_I(K^I)$
and ${\cal P}_{II}(K^{II})$.\cite{Falicov2,Migliorini}

Accordingly, the renormalization of ${\cal P}_{xy}$ is obtained as
follows, following the upper Fig.\ref{lfig2} in the direction
opposite to the arrow: (i) from the bond-moving of ${\cal P}_{xy}$
with itself, obtaining $\widetilde{{\cal P}}_1$; (ii) from the
bond-moving of ${\cal P}_z$ with itself, obtaining $\widetilde{{\cal
P}}_2$; (iii) from the decimation of $\widetilde{\cal P}_1$ and
$\widetilde{\cal P}_1$, obtaining $\widetilde{{\cal P}}_3$; (iv)
from the decimation of $\widetilde{\cal P}_2$ and $\widetilde{\cal
P}_1$, obtaining $\widetilde{{\cal P}}_4$; (v) from the decimation
of ${\cal P}_{xy}$ and ${\cal P}_{xy}$, obtaining $\widetilde{{\cal
P}}_5$; (vi) from the decimation of $\widetilde{\cal P}_3$ and
$\widetilde{\cal P}_1$, obtaining $\widetilde{{\cal P}}_6$; (vii)
from the decimation of $\widetilde{\cal P}_4$ and $\widetilde{\cal
P}_2$, obtaining $\widetilde{{\cal P}}_7$; (viii) from the
decimation of $\widetilde{\cal P}_5$ and ${\cal P}_{xy}$, obtaining
$\widetilde{{\cal P}}_8$; (ix) from the bond-moving of
$\widetilde{\cal P}_6$ and $\widetilde{\cal P}_7$, obtaining
$\widetilde{{\cal P}}_9$; (x) from the bond-moving of
$\widetilde{\cal P}_7$ and $\widetilde{\cal P}_7$, obtaining
$\widetilde{{\cal P}}_{10}$; (xi) from the bond-moving of
$\widetilde{\cal P}_9$ and $\widetilde{\cal P}_{10}$, obtaining
$\widetilde{{\cal P}}_{11}$ (xii) finally, from the bond-moving of
$\widetilde{\cal P}_{11}$ and $\widetilde{\cal P}_8$, obtaining the
renormalized quenched distribution ${\cal P}_{xy}'$.  Thus, in each
renormalization-group step, the renormalized distribution ${\cal
P}_{xy}'$ is obtained from the convolutions of 27 unrenormalized
distributions ${\cal P}_{xy}$ and ${\cal P}_z$.  The renormalized
distribution ${\cal P}_z'$ is similarly obtained from the
convolutions of 27 unrenormalized distributions ${\cal P}_{xy}$ and
${\cal P}_z$, but with a different sequencing dictated by the lower
Fig.\ref{lfig2}.

The renormalization-group transformations of the quenched
probability distributions ${\cal P}_{xy}$ and ${\cal P}_z$, given in
the preceding paragraph, are implemented numerically, resulting in a
distribution of interaction-strength values and a probability
associated with each value, namely a histogram. Thus, the initial
$\pm K^u$ double-delta distribution functions, described after
Eq.(\ref{eq:1}), are of course not conserved under the scale
coarsening of the renormalization-group transformation. The number
of histograms increases after each convolution.  When a maximum
number of histograms, set by us, is reached, a binning procedure is
applied \cite{Falicov2,Migliorini}:  Before each convolution, the
range of interaction values is divided into bins, separately for
positive and negative interactions.  The interactions falling into
the same bin are combined according to their relative probabilities.
The convolution then restores the set maximum number of histograms.
In this work, we have used the maximum number of 90,000 for
histograms for each distribution ${\cal P}_{xy}$ and ${\cal P}_z$.

\begin{table}
\begin{tabular}{|c|c|c|c|c|}
\hline \parbox{0.5in}{Phase} &
\parbox{0.6in}{$<K_+^{xy}>$} &
\parbox{0.6in}{$<K_-^{xy}>$} & \parbox{0.5in}{$<K_+^z>$} &
\parbox{0.5in}{$<K_-^z>$} \\
\hline Ferro &  $+\infty$ & 0 & $+\infty$ & 0 \\
\hline Antiferro &  0 & $-\infty$ & 0 & $-\infty$ \\
\hline Columnar &  0 & $-\infty$ & $+\infty$ & 0 \\
\hline Layered &  $+\infty$ & 0 & 0 & $-\infty$ \\
\hline Spin Glass &  $+\infty$ & $-\infty$ & $+\infty$ & $-\infty$\\
\hline Para &  0 & 0 & 0 & 0 \\
\hline
\end{tabular}
\caption{Sinks of the renormalization-group flows in the different
phases. These sinks are characterized here in terms of the average
positive and negative interactions of their limiting quenched
probability distribution.}
\end{table}

\section{Phase Diagrams and Fixed Distributions}

We have obtained the global phase diagram of the uniaxially
anisotropic $d=3$ spin-glass system in terms of the original
interactions and probabilities $(K^{xy},K^z,p_{xy},p_z)$. In each
thermodynamic phase, quenched probability distributions flow, under
repeated renormalization-group transformations, to a limiting
behavior (sink) characteristic of that thermodynamic phase. Phase
boundary points flow to their own characteristic (unstable) fixed
distributions, shown below.  Analysis at these unstable fixed
distributions yields the order of the phase
transitions.\cite{Falicov2,Migliorini}

\begin{figure}
\includegraphics*[scale=1]{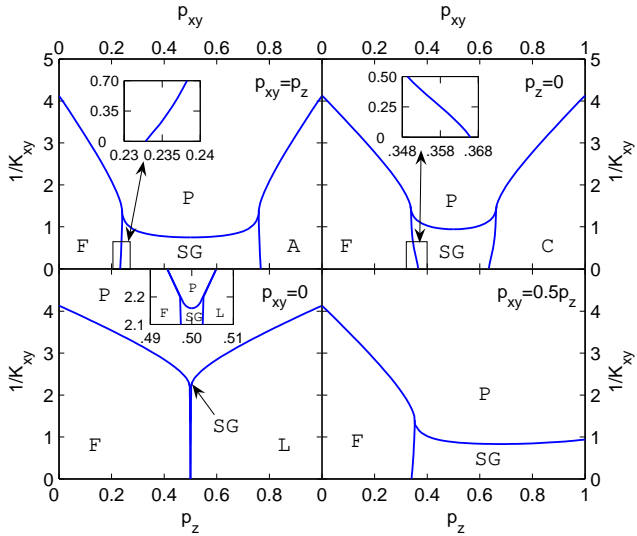}
\caption{(Color on-line) Temperature-concentration phase diagrams
for isotropically mixed (upper left), transverse (upper right),
longitudinal (lower left), and $p_{xy} = 0.5p_z$ spin-glass systems.
In all cases, $K^z/K^{xy}=0.5$.  The upper left and right phase
diagrams are seen to be, respectively, reentrant and forward, namely
with a ferromagnetic phase that, respectively, recedes from or
proceeds towards the spin-glass phase as temperature is lowered, as
clearly seen in the insets.  There are no points obeying Nishimori
symmetry in the phase diagrams of this figure.  Note the remarkably
narrow spin-glass phase, reaching zero-temperature, in the
longitudinal spin-glass system, as also seen in the inset.  All
phase transitions in this figure are second order.}\label{lfig3}
\end{figure}

\begin{figure}
\includegraphics*[scale=1]{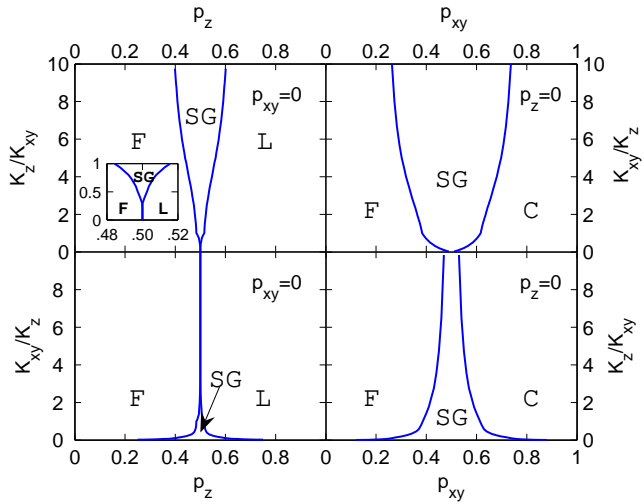}
\caption{(Color on-line) Zero-temperature phase diagrams of the
longitudinal (left column) and transverse (right column) spin-glass
systems. With the appropriate reversal in variables, the transverse
and longitudinal spin-glass phase diagrams are seen here to be
qualitatively similar, but quantitatively different.  The spin-glass
phase is more extensive in the transverse case.  All phase
transitions in this figure are second order.}\label{lfig4}
\end{figure}

We find six different phases for this system, with corresponding
sinks characterized in Table I in terms of the average positive and
negative interactions of the limiting distribution.  These phases
are the ferromagnetic, antiferromagnetic, layered, columnar,
spin-glass ordered phases and the disordered paramagnetic phase.  In
the layered phase, the spins are mutually aligned in each $xy$
plane; these planes of mutually aligned spins form an
antiferromagnetic pattern along the $z$ direction.  In the columnar
phase, the spins are mutually aligned along the $z$ direction; these
lines of mutually aligned spins form an antiferromagnetic pattern
along the $xy$ directions. Both of these phases are thus distinct
from the antiferromagnetic phase, which is antiferromagnetic in all
three directions.  There is a single spin-glass phase, extending to
anisotropic systems.

\subsection{Phase Diagrams with no Nishimori Symmetry}

Cross-sections of the global phase diagram are given in
Figs.\ref{lfig1}, \ref{lfig3}, and \ref{lfig4}.  All phase
transitions in these figures are second order. Fig.\ref{lfig1} shows
constant-temperature cross-sections of the global phase diagram as a
function of $p_{xy}$ and $p_z$.  At low temperatures (high
$K^{xy}$), the central spin-glass (SG) phase separates the corner
ferromagnetic (F), columnar (C), antiferromagnetic (A), and layered
(L) phases.  The diagrams are twofold symmetric along each axis, but
not fourfold symmetric, due to the difference between transverse
($p_z=0$) and longitudinal ($p_{xy}=0$) spin glasses. As temperature
increases, the paramagnetic (P) phase appears at the central point,
first reaches the transverse spin-glass system and eliminates the
spin-glass phase, then reaches the longitudinal spin-glass system
and eliminates the spin-glass phase. Fig.\ref{lfig3} shows
temperature-concentration phase diagrams for isotropically mixed,
transverse, longitudinal, and $p_{xy} = 0.5p_z$ spin-glass systems.
The upper left and right phase diagrams are seen to be,
respectively, reentrant and forward, namely with a ferromagnetic
phase that, respectively, recedes from or proceeds towards the
spin-glass phase as temperature is lowered, as clearly seen in the
insets.  The Nishimori symmetry (see below) is obeyed only at four
isolated ordinary points in each cross-section in Fig.\ref{lfig1}
and is not obeyed at any point in the phase diagrams in Figs.
\ref{lfig3}, \ref{lfig4}, so that the forward behavior is not
excluded by the rigorous results \cite{NishimoriA, NishimoriB}.

A remarkably narrow spin-glass phase, reaching zero-temperature,
occurs in the longitudinal spin-glass system. Zero-temperature phase
diagrams are shown in Fig.\ref{lfig4} for the longitudinal (left
column) and transverse (right column) spin-glass systems.  With the
appropriate reversal in variables, the longitudinal and transverse
spin-glass phase diagrams are seen in this figure to be
qualitatively similar, but quantitatively different. The spin-glass
phase is more extensive in the transverse case. This can be
understood from the more extensive intermixing of the ferromagnetic
and antiferromagnetic bonds.

\subsection{Temperature-Concentration Phase Diagrams with Nishimori Symmetry Curved Lines}

\begin{figure}
\centering
\includegraphics*[scale=1]{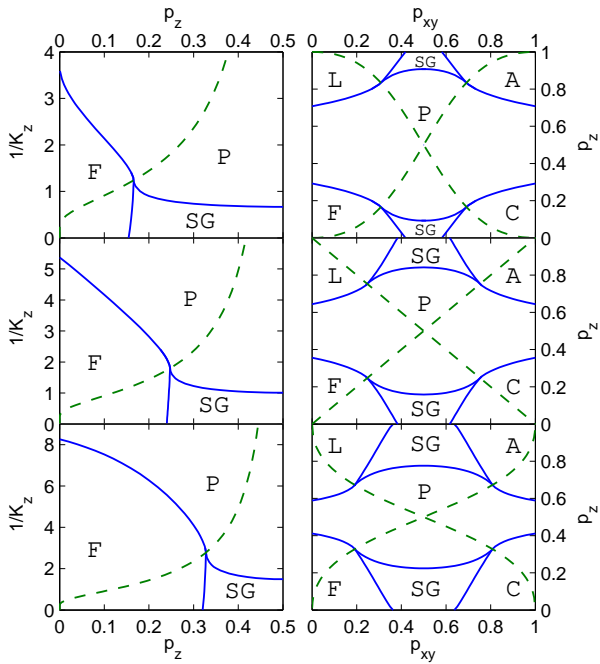}
\caption{(Color on-line) Phase diagrams with Nishimori symmetry
lines (dashed) for different anisotropy parameters: The ratio
$K^{z}/K^{xy}$ is 2, 1 and 0.5 from top to bottom.  In the left
column, $p_{xy}$ satisfies the Nishimori condition.  In the right
column, $K_z$ satisfies the Nishimori condition.  All phase
transitions in this figure are second order.}
\end{figure}

\begin{figure}[right]
\centering
\includegraphics*[scale=1]{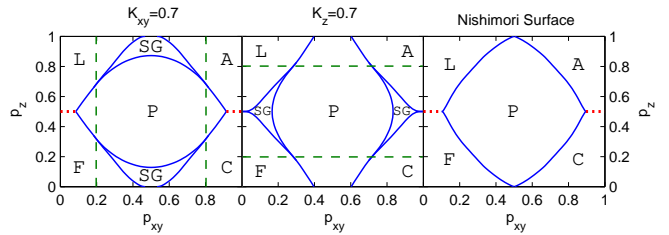}
\caption{(Color on-line) The Nishimori condition for $K_{z}$ is held
throughout the leftmost figure and for $K_{xy}$ throughout the
center figure. The complimentary Nishimori condition, for $K_{xy}$
and $K_{z}$ respectively, is held along the dashed straight lines,
which intersect the ordered (F, L, A, or C)-spinglass-paramagnetic
multicritical points. In the rightmost figure both conditions are
satisfied throughout the figure.  In this figure, the phase
boundaries around the paramagnetic phase are actually lines of the
multicritical points where the paramagnetic, ordered (F, L, A, or
C), and spin-glass (not seen in this cross-section) phases meet. In
the side figures, first-order boundaries (dotted) occur between the
ferromagnetic and layered phases, and between the antiferromagnetic
and columnar phases, terminating at $d = 2$ critical points.  All
other phase transitions (full lines) in this figure are second
order.}
\end{figure}

\begin{figure}[right]
\centering
\includegraphics*[scale=1]{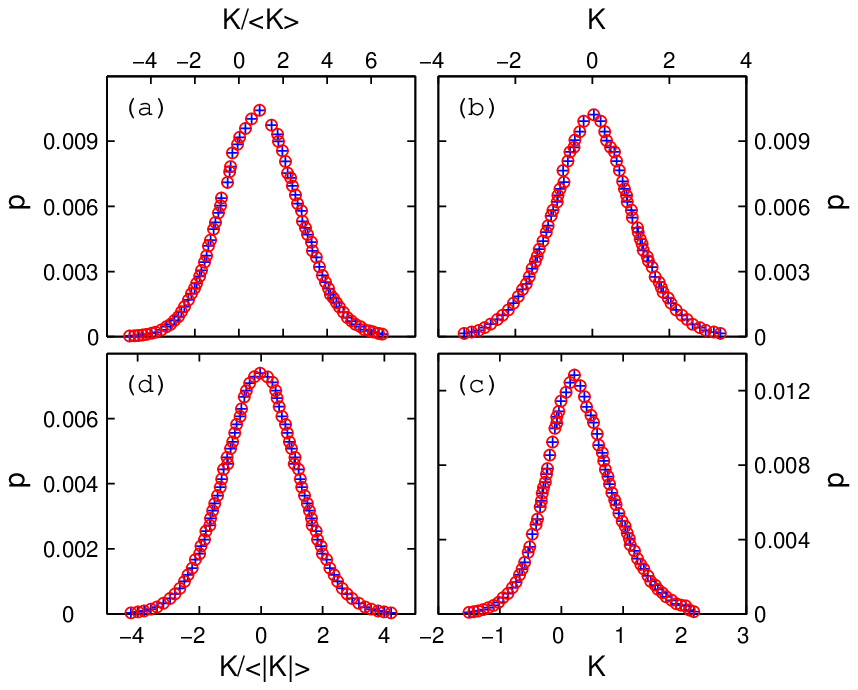}
\caption{(Color on-line) Fixed distributions, with circles and
crosses showing one renormalization-group transformation and thereby
by their exact superposition attesting to the fixed nature of the
distributions. The distributions have been binned for exhibition
purposes.  (a) For the ferromagnetic-spinglass phase boundary, a
runaway to infinite coupling; (b) for the paramagnetic-spinglass
phase boundary.  Both of these fixed distributions are spatially
isotropic, attracting isotropic and anisotropic boundaries, and do
not obey Nishimori symmetry.  (c) For the
ferromagnetic-spinglass-paramagnetic multicritical point.  This
fixed distribution is spatially isotropic and obeys Nishimori
symmetry.  This fixed distribution attracts the isotropic
multicritical point, which obeys Nishimori symmetry, and anisotropic
multicritical points, which obey and do not obey Nishimori symmetry.
The fixed distributions for the antiferromagnetic-spinglass,
columnar-spinglass, layered-spinglass phase boundaries and for the
antiferromagnetic-spinglass-paramagnetic,
columnar-spinglass-paramagnetic, layered-spinglass-paramagnetic
multicritical points are as shown here in (a) and (c) respectively,
but with the appropriate $K_{xy} \rightarrow -K_{xy}$ and/or $K_z
\rightarrow -K_z$ reflections.  (d) Fixed distribution for the
spin-glass phase. This phase sink is an isotropic runaway,
attracting both spatially isotropic and anisotropic spin-glass phase
points, and does not obey Nishimori symmetry.}
\end{figure}

The Nishimori symmetry condition \cite{NishimoriA, NishimoriB} for
isotropic systems,

\begin{equation}\label{eq:5}
\frac{1-p}{p}=e^{\pm 2K},
\end{equation}

\noindent generalizes, for uniaxially anisotropic spin-glass
systems, to

\begin{equation}\label{eq:6}
\frac{1-p_{xy}}{p_{xy}}=e^{\pm 2K_{xy}}\quad \text{and} \quad
\frac{1-p_{z}}{p_{z}}=e^{\pm 2K_{z}}.
\end{equation}

\noindent For Nishimori symmetry to obtain, both equations have to
be satisfied, but the signs in the exponents can be chosen
independently. The Nishimori condition, in its general form

\begin{equation}\label{eq:6b}
\frac{{\cal P}_u (-K_u)}{{\cal P}_u (K_u)}=e^{\pm 2K_u}
\end{equation}

\noindent for each histogram pair of each distribution, is invariant
(closed) under our renormalization-group transformation.

If one of the two conditions in Eq.(\ref{eq:6}) is fixed, phase
diagram cross-sections are obtained, in which Nishimori symmetry
holds along a line.  Thus, throughout the three phase diagrams on
the left in Fig.5, the condition on $(K_{xy},p_{xy})$ is fixed.  The
condition on $(K_z,p_z)$, and therefore Nishimori symmetry, is
satisfied along the dashed lines on the left in Fig.5. In these
temperature versus concentration phase diagrams, it is seen that the
multicritical points between the ferromagnetic, spin-glass, and
paramagnetic phases lie on the Nishimori symmetry line. Furthermore,
it has been proven \cite{NishimoriA, NishimoriB} that a forward
phase diagram cannot occur below such a multicritical point that is
on the symmetry line. On the left in Fig.5, this is indeed the case,
with reentrant phase diagrams, as also seen in isotropic spin
glasses \cite{Migliorini, Nobre}.  Recall that in Sec.IVA,
multicritical points, between the same phases as here, that do not
lie on Nishimori symmetry occur with both reentrant and forward
phase diagrams.  However, the latter non-symmetric multicritical
points flow, under renormalization-group transformations, to the
(doubly unstable) fixed distribution of the symmetric multicritical
points, therefore being in the same universality class and having
the same critical exponents.

In the three phase diagrams on the right of Fig.5, the condition on
$(K_z,p_z)$ is fixed.  In these concentration-concentration phase
diagrams, the multicritical points between the ordered
(ferromagnetic, antiferromagnetic, layered, or columnar),
spin-glass, and paramagnetic phases again lie on the Nishimori
symmetry lines.

\subsection{Concentration-Concentration Phase Diagrams with Nishimori Symmetry Straight Lines}
  In the phase diagrams in Fig.5, the ratio
$K_z/K_{xy}$ is held constant.  On the left and center of Fig.6,
again the condition in Eq.(\ref{eq:6}) on one interaction is fixed
and the other interaction strength is held constant.  Thus, the
Nishimori symmetry lines becomes straight lines.  The multicritical
points between the ordered (ferromagnetic, antiferromagnetic,
layered, or columnar), spin-glass, and paramagnetic phases again lie
on the Nishimori symmetry lines.  In the left phase diagram, due to
the enforced Nishimori symmetry condition, $K_z = 0$ along the line
$p_z = 0.5$ and the system reduces to $d=2$.  Along this line,
first-order transitions between ferromagnetic and layered phases and
between antiferromagnetic and columnar phases terminate at $d=2$
critical points.  From $p_z \neq 0.5$, $d = 3$ second-order
boundaries between each ordered phase and the paramagnetic phase
terminate on the $d = 2$ critical points.  In the center phase
diagram, due to the enforced Nishimori symmetry condition, $K_{xy} =
0$ along the line $p_{xy} = 0.5$ and the system reduces to $d=1$.
Accordingly, the system is disordered (paramagnetic) along the
entire length of this line.

\subsection{The Phase Diagram Entirely Imbedded in Nishimori
Symmetry}

In the rightmost Fig.6, both conditions of Eq.(\ref{eq:6}) are
satisfied throughout the figure.  With two symmetry constraints,
this is a unique surface in the global phase diagram of our model.
The system reduces to $d = 2$ and $d =1$, as explained above, for
$p_z = 0.5$ and $p_{xy} = 0.5$ respectively.  The phase boundaries
around the paramagnetic phases are actually lines of the
multicritical points where the paramagnetic, ordered (ferromagnetic,
layered, antiferromagnetic, or columnar), and spin-glass (not seen
in this cross-section) phases meet.

No spin-glass phase occurs within the Nishimori-symmetric subspace.
The phase transitions seen in the rightmost Fig.6, namely
ordered-spinglass-paramagnetic multicritical and
ferromagnetic-layered, antiferromagnetic-layered first-order
transitions, are the only phase transitions of the system that occur
under Nishimori symmetry.

\subsection{Fixed Distributions}

The fixed distributions underpinning the phase diagrams of this
system are given in Fig.7.  The fixed distributions for the
ferromagnetic-spinglass boundary, paramagnetic-spinglass boundary,
and the ferromagnetic-spinglass-paramagnetic multicritical points
are spatially isotropic, but attract both spatially isotropic and
anisotropic phase transitions.  The fixed distribution for the
ferromagnetic-spinglass-paramagnetic multicritical points obeys
Nishimori symmetry, but attracts multicritical points that obey and
do not obey Nishimori symmetry.  In the latter cases, as seen above,
both reentrant and forward phase diagrams occur.

The fixed distributions for the antiferromagnetic-spinglass,
columnar-spinglass, layered-spinglass phase boundaries and for the
antiferromagnetic-spinglass-paramagnetic,
columnar-spinglass-paramagnetic, layered-spinglass-paramagnetic
multicritical points are as shown in Fig.7 (a) and (c) respectively,
but with the appropriate $K_{xy} \rightarrow -K_{xy}$ and/or $K_z
\rightarrow -K_z$ reflections.

\section{Conclusion}

The exact solution of the spatially uniaxially anisotropic spin
glass on a $d=3$ hierarchical lattice yields new phase diagrams. In
view of the semiquantitative agreement between spatially isotropic
spin-glass results on cubic and hierarchical lattices
\cite{Migliorini}, it would certainly be worthwhile to investigate
on cubic lattices the new phenomena found in the present study.
Furthermore, the exact study of spin glasses on fully anisotropic
$d=3$ hierarchical lattices \cite{Erbas} may yield even more new
phase transition phenomena.

\begin{acknowledgments}
This research was supported by the Scientific and Technological
Research Council (T\"UB\.ITAK) and by the Academy of Sciences of
Turkey.
\end{acknowledgments}

\end{document}